\documentclass[conference, a4paper]{IEEEtran}
\usepackage{graphicx,xcolor,tikz}
\usepackage{comment}
\usepackage{float}
\usepackage{multicol}
\usepackage{cite}
\usepackage{makecell}
\usepackage{siunitx}
\usepackage{multirow}
\usepackage{mathtools}
\usepackage{amsmath,amssymb, amsfonts}
\usepackage{tabulary}
\usepackage{stfloats}
\usepackage{cite}
\usepackage[caption=false]{subfig}
\usepackage{mathtools}
\usepackage{balance}
\usepackage[normalem]{ulem}
\def\nb0{{\mathbf{0}}}
\def\nb1{{\mathbf{1}}}

\usepackage{tikz,epstopdf}

\usetikzlibrary{arrows}
\usepackage{lipsum,graphicx}
\usepackage[para]{threeparttable}
\usepackage{epsfig,graphics,graphicx,subfig,amssymb,amstext,amsmath,algorithm, algorithmic,wrapfig,multirow}
\usepackage{adjustbox}
\usepackage{siunitx}
\usepackage{xcolor}
\usepackage{breakurl}
\usepackage[breaklinks=true]{hyperref}
\usetikzlibrary{chains,arrows,calc ,positioning}
\usepackage{graphicx}
\usepackage{colortbl, xcolor}
\usepackage{enumitem}
\usepackage{url}
\usepackage[utf8]{inputenc}
\usepackage{cuted}
\usepackage{stfloats}
\setitemize{noitemsep,topsep=0pt,parsep=0pt,partopsep=0pt}
\def\BibTeX{{\rm B\kern-.05em{\sc i\kern-.025em b}\kern-.08em T\kern-.1667em\lower.7ex\hbox{E}\kern-.125emX}}

\usepackage[left=1.62cm,right=1.62cm, top=0.75in, bottom = 1.05in]{geometry}
\newcommand{\Ps}{P_{\text{S}}}

\newcommand{\PISI}{P_{\text{ISI}}}

\newcommand{\PIMP}{\bar{P}_{\text{I}}^{\text{M}}}
\newcommand{\PIDP}{\bar{P}_{\text{I}}^{\text{D}}}

\newcommand{\YICI}{Y_{\text{ICI}}}
\newcommand{\YS}{Y_{\text{S}}}
\newcommand{\YISI}{Y_{\text{ISI}}}
\newcommand{\Ncp}{N_{\text{cp}}}

\let\oldthebibliography\thebibliography
\renewcommand{\thebibliography}[1]{%
  \oldthebibliography{#1}%
  \setlength{\itemsep}{0pt}%
  \setlength{\parsep}{0pt}%
  \setlength{\parskip}{0pt}%
}

\IEEEoverridecommandlockouts

\begin{document}
%\title{Doppler Effect Analysis in OFDM-Based Sub-THz LEO Satellite Systems}
%\title{Impact of Doppler Shift on Sub Terahertz\\Low Earth Orbit Satellite Communication\vspace{-3mm}}
%\title{Sub Terahertz LEO Satellite Communication:\\Does the Doppler Shift Matter?\vspace{-3mm}}
\title{Sub Terahertz LEO Satellite Communication:\\Is Doppler Shift a Major Challenge?\vspace{-3mm}}

\author{
\IEEEauthorblockN{Seongjoon Kang$^{\dagger}$, Parisa Ramezani$^{\flat}$, Ki Won Sung$^{\dagger}$, and Vitaly Petrov$^{\dagger}$}
\IEEEauthorblockA{$^{\dagger}$KTH Royal Institute of Technology, Stockholm, Sweden, (\{skang, sungkw, vitalyp\}@kth.se)}
\IEEEauthorblockA{$^{\flat}$ University of Technology Sydney, Sydney, Australia, (parisa.ramezani@uts.edu.au)\vspace{-2mm}
%\thanks{ \textcolor{black}{V. Petrov, K.W. Sung, and S. Kang were supported by some grants from KTH. The work of P. Ramezani was supported by the Academy of Australia. Add more senteces here~!}}
}}

\maketitle

\begin{abstract}
In this work, we investigate the impact of Doppler frequency shift on sub-terahertz low Earth orbit (LEO) satellite communications using OFDM transmission. To comprehensively evaluate the severity of the Doppler shift impact on sub-terahertz satellite-to-ground links, we first put together an evaluation framework that jointly accounts for major Doppler-induced and non-Doppler-induced signal impairments in such systems. We then compare and contrast three possible engineering approaches to deal with Doppler shift in such a communication system. Our study concludes that (i)~while just ignoring the Doppler effect in such high-frequency and high-mobility systems substantially compromises the performance and reliability, (ii)~most of the negative effects can be eliminated by implementing only a small set of low-complexity measures. We particularly show that even a single Doppler shift estimation over a 5-min-long satellite pass over the ground user location may be sufficient to ensure high performance throughout the entire connectivity period. Hence, while Doppler shift is an important effect to account for in sub-terahertz LEO satellite communications, it should not constitute a critical design challenge for these emerging systems.
\end{abstract}


%%%%%%%%%%%%%%%%%%%%%%%%%%%%%%%%%%%%%%%%%%%%%%%%%%%%%%%%%%%%%%%%%%%%%%%%%%%%%%%%%%%%%%%%%%%%
\section{Introduction}
\label{sec:intro}
%In the next era of communications, low Earth orbit (LEO) satellites are to provide fiber-like global connectivity worldwide, enabling ubiquitous Internet access for users on the ground, at sea, and in the air~\cite{giordani2020non, lin2022path}. 
\textcolor{black}{In the next era of communications, low Earth orbit (LEO) satellites are envisioned to deliver broadband connectivity worldwide, enabling ubiquitous Internet access for users on the ground, at sea, and in the air~\cite{giordani2020non, lin2022path}}.
%\textcolor{red}{[Reviewer 2] The very first statement must be at least smoothened. Performance of LEO systems in terms of throughput are significantly lower than optical fibre systems.}
To satisfy the continuously growing requirements, microwave and millimeter-wave radios that are prevalently used in today's LEO satellite constellations are expected to be eventually complemented by the ones featuring wider frequency bands at higher frequencies, including the sub-terahertz (sub-THz) spectrum above 100~GHz~\cite{katwe2025cmwave,chong_space_sagin}.

Meanwhile, the landscape of sub-THz radios has also evolved substantially over the recent three decades. What started back in the 20th century with pioneer scientific explorations of exclusively short-range connectivity~\cite{60016} has eventually evolved into a heterogeneous technological setup targeting also mid-range (e.g., next-generation Wi-Fi or cellular services~\cite{9838922}) or even long-range connectivity scenarios (such as stationary backhaul links over $\SI{1}{\kilo \meter}$ long~\cite{sen2023multi} and airborne sub-THz links~\cite{11538050}). Importantly, the academic effort in this field is also complemented by recent industry demos~\cite{mobile_thz_trial_japan}, regulatory work~\cite{mike_thz_spectrum}, and the appearance of the first global standards using these new frequency bands~\cite{ieee_802_15_3d_standard_2024}.

With airplane-to-airplane sub-THz wireless link successfully demonstrated~\cite{afrl_flight_demo_thz}, sub-THz connectivity between a ground station and a LEO satellite naturally becomes the next research frontier~\cite{katwe2025cmwave,chong_space_sagin}. Here, in parallel to ongoing theoretical studies, several groups worldwide are working on \emph{prototyping such a system within the next few years}~\cite{sat_thz_prototype}. However, multiple challenges arise on the way from theoretically exploring sub-THz LEO satellite communication systems toward successfully building them in hardware. These include, among others, power, efficiency, and weight constraints~\cite{Aliaga2025Payload}, the design of reliable beam realignment procedures~\cite{11309975}, as well as modeling and compensating for the possible adverse weather conditions~\cite{ma2018effect,Aliaga2025Scintillation}.

One major distinct effect here, giving the motivation for the current study, is \emph{the Doppler frequency shift} -- a change in the wireless signal due to the relative mobility of the nodes. From the fundamental physics, the Doppler shift grows fast with \emph{both} the signal frequency and the nodes' relative velocity. Hence, for sub-THz LEO satellite-to-ground links (with 100~GHz+ frequency and e.g., $\approx$\SI{7.56}{\km / \second} satellite speed at an altitude of \SI{600}{\km}), \emph{the impact of Doppler effects might be substantial, possibly even challenging the feasibility of a reliable sub-THz satellite link.}

The present work is built upon extensive prior studies of Doppler shift in LEO satellite communication systems. This includes one of the early works by I.~Ali et al.~\cite{Ali1998Doppler} deriving the fundamentals for the Doppler shift characterization in LEO-centric scenario geometries. To be consistent with the 3GPP decision to adopt OFDM as a primary waveform for 5G+ and 6G-grade access links~\cite{6G_OFDM_agreement}, including non-terrestrial network (NTN) satellite-to-ground links~\cite{3GPP38300}, we partially reference the results from~\cite{zuo2023ofdm} and~\cite{huang2021synchronization} that investigated the performance of an OFDM-based LEO satellite link.

Various other contributions explored neighbouring subjects, including but not limited to \cite{na2013channel} that analyzes the throughput of LEO OFDM satellite systems using a measurement-based channel model without analytically deriving the impact of Doppler shift on subcarrier orthogonality. Recently, the study of the residual Doppler shift impact on the coverage in LEO OFDM systems was presented~\cite{tanash2026performance}, but its analysis is purely geometric and considers only two fixed subcarrier spacings.

Hence, despite substantial prior work on the Doppler effect on LEO satellite communication in general, \emph{to the best of the authors' knowledge, no prior work has jointly characterized the superposition of all the major Doppler-induced and non-Doppler-induced impairments of the sub-THz signal over a LEO satellite link as functions of the number of subcarriers}, namely the Doppler-induced inter-carrier interference (ICI), the multipath-induced inter-symbol interference (ISI), the Doppler-induced loss in the desired signal power, and accumulating Doppler-induced phase error.

To address the gap, in this work, we present a link- and system-level analysis of the abovementioned Doppler-induced impairments in OFDM-based sub-THz LEO satellite communications, jointly modeling the ICI and the ISI levels. Based on this analysis, we identify Doppler-aware OFDM subcarrier configurations to maximize the performance of sub-THz LEO satellite-to-ground links. We finally compare different implementation options (with various levels of ``Doppler-awareness'') to quantify the implications of (i)~ignoring, (ii)~partially, or (iii)~fully compensating for the Doppler shift in sub-THz LEO satellite communication.

Our analysis concludes that, while the Doppler shift has a major effect at sub-THz frequencies, for the proposed low-complexity Doppler compensation scheme (referred to as Configuration~$3$), \emph{as few as one Doppler shift estimation} often becomes sufficient to ensure high-throughput reliable data exchange between the nodes during the entire connectivity period. The study thus serves as a useful input toward prototyping such systems~\cite{sat_thz_prototype}, as well as evaluating more complex sub-THz LEO satellite connectivity scenarios with multiple ground users and/or a full-scale LEO satellite constellation.

%The rest of the paper is organized as follows. The scenario and key modeling assumptions are outlined in Section~\ref{sec:system_model}. In Section~\ref{sec:impairments}, the evaluation framework is presented, and the key performance metrics are derived mathematically. Numerical results are introduced and discussed in Section~\ref{sec:numerical_results}. Conclusions are drawn in the last section.

%%%%%%%%%%%%%%%%%%%%%%%%%%%%%%%%%%%%%%%%%%%%%%%%%%%%%%%%%%%%%%%%%%%%%%%%%%%%%%%%%%%%%%%%%%%%
\section{System Model}
\label{sec:system_model}

%%%%%%%%%%%%%%%%%%%%%%%%%%%%%%
\subsection{Scenario Geometry}
\label{subsec:scenario}
\begin{figure}[!t]
    \centering
    \includegraphics[width =0.99\columnwidth]{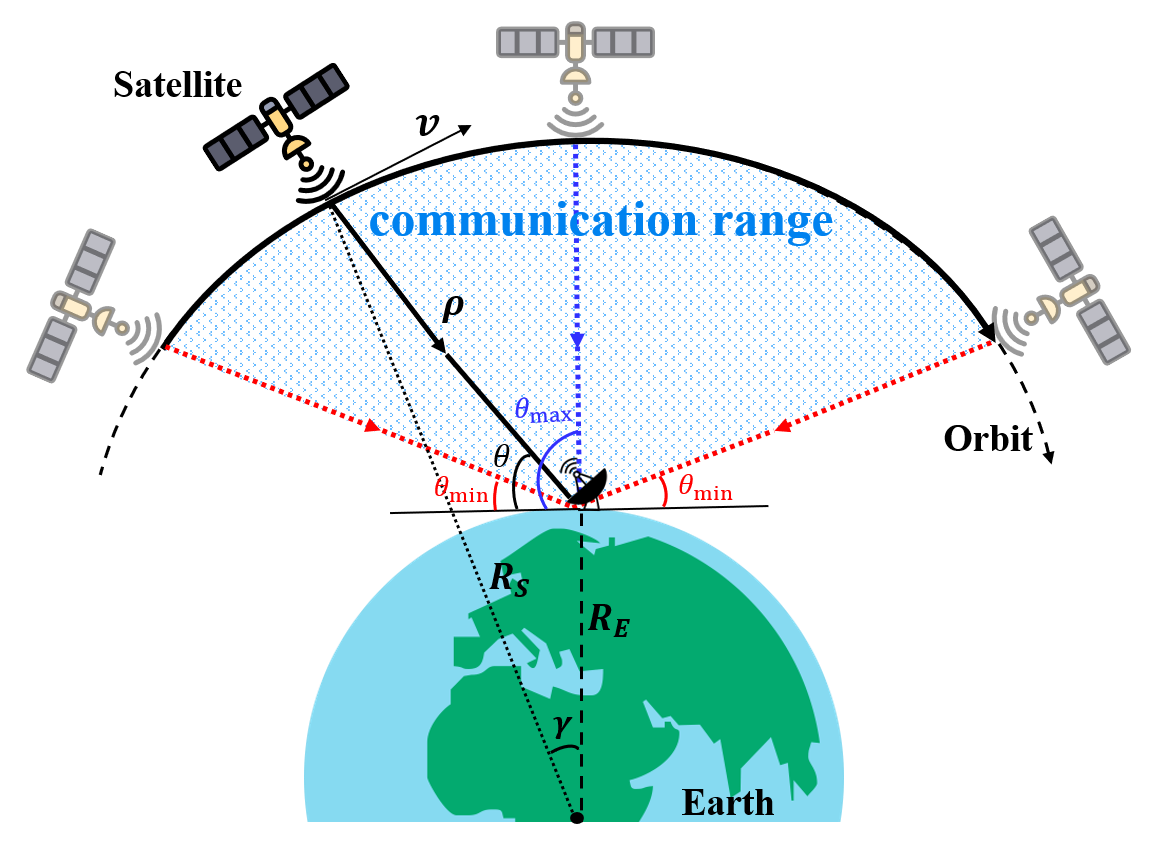} 
    \vspace{-2mm}
    \caption{Evaluated satellite-to-ground sub-THz communication system.}
    \vspace{-5mm}
    \label{fig:Earth_and_groundUser}
\end{figure}

We model a sub-THz wireless link between a LEO satellite at an altitude $h_\text{S}$ and a ground user (GU), as in Fig.~\ref{fig:Earth_and_groundUser}.
For notational clarity, we evaluate a satellite-to-GU downlink, while most of the mathematical analysis is, in principle, reciprocal and thus also applicable to the uplink scenario. We follow a circular orbit assumption; hence, eccentricity, $e$, is approximated as zero~\cite{circular_orbit_table}\footnote{Many practical LEO orbits feature $e \leq 10^{-3}$ (e.g., \cite{circular_orbit_table}, Table~II).}. The actual communication range between the nodes is limited to the practical set of elevation angles $\theta \geq \theta_{\min}$ along the satellite trajectory (see Fig.~\ref{fig:Earth_and_groundUser}). The Earth radius is considered constant along the path and denoted by $R_\text{E}$.

%%%%%%%%%%%%%%%%%%%%%%%%%%%%%%
\subsection{Channel Model}
\label{subsec:channel_model}
The large-scale channel gain mainly consists of the free space path loss (FSPL) and shadow fading, since the clutter loss is negligible in line-of-sight (LOS) channels~\cite{3GPP38811}.
  The FSPL $P_{\text{L}}$ is modeled as
\begin{equation}
   \label{eq:pathloss}
    P_\text{L} = \frac{L_\text{A}}{G_\text{T} G_\text{R}} \left( \frac{4\pi\rho f_\text{c}}{c}\right)^2,
\end{equation}
where $G_\text{T}$ and $G_\text{R}$ are the transmit and receive antenna gains, $\rho$ is the distance between the satellite and the GU, and  $L_\text{A}$ accounts for additional propagation losses, such as atmospheric absorption and rain attenuation~\cite{3GPP38811}. 
%\textcolor{red}{[Reviewer 2] Section II-B The free space losses usually doesn’t include rain attenuation and other tropospheric impairments. Free space losses are deterministic while rain attenuation and other tropospheric impairments are stochastic processes.}
In this study, we specifically isolate the Doppler-related effects from the impact of imperfections in dynamic beam realignment (that vary drastically depending on the selected realignment algorithm and protocols~\cite{mobile_thz_trial_japan,truly_mobile_thz}). For this purpose, we assume perfect beam alignment between the satellite and the ground user, i.e., both $G_\text{T}$ and $G_{\text{R}}$ are stationary values.% \textcolor{black}%{Following~\cite{maral2020satellite}, the maximum antenna gain can be obtained by assuming a circular antenna aperture, as follows:
%\begin{equation}
%G_{\max} = \eta \left(\frac{\pi D}{\lambda}\right)^2,
%\end{equation}
%where $\eta$ denotes the aperture efficiency, $D$ denotes the antenna diameter, and $\lambda$ denotes the wavelength of the transmitted signal.}
%\textcolor{red}{[Reviewer 1] It is recommended to add the antenna type and explain the max antenna gain, i.e., we should use typical antenna gain instead of max as a practical assumption. }

\textcolor{black}{We assume an \emph{LOS-dominant} link between the satellite and the GU, motivated by the fact that LEO satellite systems are primarily envisioned to extend connectivity to under-served areas~\cite{giordani2020non, lin2022path} such as rural areas.}
Thus, following a common modeling approach for similar scenarios~\cite{hassan2024spaceris, bakhsh2025direct}, the communication channel is modeled as a frequency-selective Rician fading channel consisting of one dominant line-of-sight (LOS) path and multiple non-LOS (NLOS) scattered paths. Due to the strong LOS propagation characteristic of satellite communications, the first tap is assumed to contain the LOS component, and the channel coefficient $h[l]$ of the $l$th tap is modeled as
\begin{equation}
h[l] = \frac{1}{\sqrt{P_{\mathrm{L}}}} \left( \sqrt{\frac{K}{K+1}}\, e^{j\phi_0}\, \delta[l] 
       + \sqrt{\frac{1}{L(K+1)}}\, \tilde{h}[l] \right),
\end{equation}
%h[l]
%= \frac{1}{\sqrt{P_\text{L}}}
%&\sqrt{\frac{K}{K+1}}~e^{j\phi_0}~\delta[l] \notag\\
%&\quad + \frac{1}{\sqrt{P_\text{L}}}
%\sum_{m=0}^{L-1}
%\sqrt{\frac{1}{L(K+1)}}~\tilde{h}[m]~\delta[l-m]
where $l = 0, \ldots, L-1$ denotes the tap index, $\delta$ is the Kronecker delta function, $L$ denotes the length of multipath channel $\boldsymbol{h} = [h[0], h[1], ... , h[L-1]]$,  \(\phi_0\) is the LOS phase, $K$ denotes the Rician $K$-factor, and $\tilde{h}[l] \sim \mathcal{CN}(0,1)$ represents the scattered component of the NLOS path. Throughout this paper, the uncorrelated scattering assumption is adopted for analytical tractability~\cite{proakis2001digital}.

%Building upon the 3GPP considerations for the NR non-terrestrial networks~\cite{3GPP38300, 3gpp38211}, we also primarily focus on the OFDM signals.
Building upon the 3GPP considerations for the NR non-terrestrial 
networks~\cite{3GPP38300, 3gpp38211}, this work primarily focuses on OFDM transmission under the assumption of perfect symbol timing synchronization. The transmit power is set to a constant $P_{\text{T}}$, equally allocated across bandwidth $B$ and $N$ OFDM subcarriers, so the subcarrier spacing is $\Delta f = B/N$ and the TX power per subcarrier $m$ becomes $P_m = P_\text{T}/N$.

%%%%%%%%%%%%%%%%%%%%%%%%%%%%%%%%%%%%%%%%%%%%%%%%%%%%%%%%%%%%%%%%%%%%%%%%%%%%%%%%%%%%%%%%%%%%%%%%%%%%%%%%%%%%%%%%%%%%%%%%%%%%%%%%%%%%%%%%%%%%%%%%%%%%%%

%%%%%%%%%%%%%%%%%%%%%%%%%%%%%%%%%%%%%%%%%%%%%%%%%%%%%%%%%%%%%%%%%%%%%%%%%%%%%%%%%%%%%%%%%%%%
\section{Analysis of Doppler-related Impairments}
\label{sec:impairments}
In this section, we analyze the OFDM transmission impairments over a satellite-to-ground sub-THz wireless link induced by multipath propagation and Doppler shift. 
%We start by recalling the derivation of the Doppler shift itself for the given scenario in Section~\ref{subsec:doppler_shift}. We then discuss the major signal impairments introduced by the presence of Doppler shift in Section~\ref{subsec:doppler_effect_analysis}, namely, the ICI, the degradation of the desired signal power over a given subcarrier, and the accumulating phase offset. We later corroborate the analysis in Section~\ref{subsec:isi_analysis} to also account for multipath propagation effects leading to ISI. Finally, all the individual effects are combined together in Section~\ref{subsec:kpi_analysis} to derive the key performance indicators (KPIs) of interest, including the average signal-to-interference-plus-noise ratio (SINR), the average bit error rate, and the average effective throughput.
%\textcolor{red}{[Reviewer 2] Section III The first sentences are useless. They can be removed. The preliminary summary of the contents of the section just makes the paper longer without adding value. }

%%%%%%%%%%%%%%%%%%%%%%%%%%%%%%
\subsection{Recap of Doppler Shift in LEO Satellite Communications}
\label{subsec:doppler_shift}
To evaluate the implications of the Doppler effect on sub-THz LEO satellite link performance, we first recap the derivation of the Doppler shift in our target scenario~\cite{Ali1998Doppler}. Recall that the general time-varying expression of the Doppler shift $f_\text{d}$ is
\textcolor{black}{
\begin{equation}
f_\text{d}=-\frac{f_\text{c}}{c}\frac{d\rho}{dt} =\frac{f_\text{c}}{c}v_r,
\label{eq:doppler_def}
\end{equation}}
where $v_r$ is the radial velocity, $f_\text{c}$ is the carrier frequency, $c$ is the speed of light, and $\rho$ is the slant distance between the satellite and the GU. 

The elevation angle $\theta$ and the geocentric angle $\gamma$ are related through the geometric expression
\begin{equation}
\gamma=\arccos\!\Big(\frac{R_{\rm E}}{R_{\rm S}}\cos\!\big(\theta\big)\Big)-\theta,
\label{eq:gamma_theta}
\end{equation}
where $R_{\rm S} = R_{\rm E} + h_{\text{S}}$ is the distance between the satellite and the Earth center, and the slant distance $\rho(\gamma)$ is given by 
\begin{equation}
    \rho (\gamma) =\sqrt{R_{\text{E}}^2+R_{\rm S}^2-2R_{\rm E} R_{\rm S}\cos\!\big(\gamma\big)}.
    \label{eq:slant_range}
\end{equation}
%Let $\omega$ the satellite angular velocity, which is approximated as $\omega \approx  \sqrt{\frac{\mu}{R_{\rm S}^3}}-\omega_{\rm E} \cos i$, where $\mu$ is the Earth's gravitational parameter, given as $\SI{398600.4418}{\kilo\metre^3\per\second^2}$, $\omega_E$ is the Earth's rotational angular velocity, given as $\SI{7.2921159e-5}{\radian\per\second}$, and $i$ is the orbit inclination angle.
Let $\omega$ be the satellite angular velocity, which is approximated as $\omega \approx \sqrt{\frac{\mu}{R_{\rm S}^3}} - \omega_{\rm E} \cos i_{\rm o}$, where $\mu$ is the Earth's gravitational parameter, $\SI{398600.4418}{\kilo\metre\cubed\per\second\squared}$; $\omega_{\rm E}$ is the Earth's rotational angular velocity, $\SI{7.2921159e-5}{\radian\per\second}$; and $i_{\rm o}$ is the orbit inclination angle.
Assuming a zenith pass, i.e., $\theta_{\max}=90^\circ$, the radial velocity $v_r$ is given by~\cite{Ali1998Doppler}\footnote{We particularly focus on the case where the LEO satellite orbit passes \emph{exactly over} the GU; hence, the Doppler-related effects are the largest. The Doppler shift grows with the relative velocity between the nodes, which is maximized at a minimum elevation angle~\cite{Ali1998Doppler}. The Doppler-related effects in other scenarios will hence not exceed those observed in this study, which serves the aim of this work.}
\begin{equation}
    v_r = \pm\frac{R_{\rm E} R_{\rm S}\,\omega}{\rho(\gamma)} \sin (\gamma).
    \label{eq:radial_velocity}
\end{equation}
  The `$+$' sign of $v_r$ corresponds to one half of the pass (i.e., approaching), while the `$-$' sign corresponds to the other half (i.e., receding). By substituting \eqref{eq:radial_velocity} into \eqref{eq:doppler_def}, the closed-form expression for the Doppler frequency as a function of the geocentric angle $\gamma$ is given by
\begin{equation}
    f_\text{d}(\gamma) = \pm
    \frac{f_\text{c}}{c}\frac{R_{\rm E} R_{\rm S}\,\omega}{\rho(\gamma)}\
    \sin (\gamma).
    \label{eq:doppler_gamma}
\end{equation}
Note that as the elevation angle $\theta$ approaches $\theta_{\max} = 90^\circ$, the geocentric angle $\gamma$ approaches zero, yielding $f_\text{d}(\gamma) \to 0$.

%%%%%%%%%%%%%%%%%%%%%%%%%%%%%%
\subsection{Doppler-Related Effects in OFDM Systems}
\label{subsec:doppler_effect_analysis}
Let \(X(k)\) be the modulated symbol (e.g., a QPSK or a QAM symbol) on the \(k\)-th subcarrier, and $T_s$ denote the sampling time. After the inverse fast Fourier transform (IFFT), the time-domain transmission signal sampled at $n T_s$ is given by
\begin{equation}
\label {eq:modul}
x[n] = \frac{1}{\sqrt{N}} \sum_{k=0}^{N-1} X(k) e^{j \frac{2 \pi k n}{N}}, 
\quad n = 0, 1, \dots, N-1, \nonumber
\end{equation}
where the data symbols $X(k)$ are assumed to be independent and identically distributed (i.i.d.) with zero mean and unit average power, i.e., $\mathbb{E}\left[X(k)X^*(m)\right] = \delta(k-m)$. The cyclic prefix (CP) is added to the OFDM symbol by copying the last $N_{\mathrm{cp}}$ samples of $x[n]$, i.e., $x[-p] = x[N-p]$ for $p = 1, \ldots, N_{\mathrm{cp}}$. 
%and prepending  them to the OFDM symbol.
%The cyclic prefix (CP) is then formed by copying the last $N_{\mathrm{cp}}$ samples of $x[n]$ and prepending them to the OFDM symbol.
%It follows that the time domain signal $x[n]$ in \eqref{eq:modul} satisfies $\mathbb{E}\left[x[n]x^*[m]\right] = \delta[n-m]$. 
%After being  transmitted through the channel $\boldsymbol{h}$ in the presence of Doppler effects, the received signal is written as:
The signal received through $\boldsymbol{h}$ with the Doppler shift is given by
\begin{align}
\label{eq:recvd}
y[n] &=  \sum_{l=0}^{L-1} h[l] x[n-l]~e^{j2\pi n f_{d} T_s}+ y_{\text{ISI}}[n] + w[n] \nonumber \\
           &=   \sum_{l=0}^{L-1} h[l] \left( \frac{1}{\sqrt{N}} \sum_{k=0}^{N -1} X(k) e^{j \frac{2 \pi k (n-l)}{N}} \right)~e^{j2\pi n f_{d}  T_s} \nonumber \\
           &+ y_{\mathrm{ISI}}[n] + w[n], \quad n = -\Ncp,  \dots, N-1, 
\end{align}
where $w[n]$ is the noise signal, $f_\text{d}$ is the common Doppler frequency shift across multipath components, and $y_{\text{ISI}}$ is the received ISI.
Since the altitude of a LEO satellite (typically above \SI{600}{km}) is orders of magnitude larger than the distances to local ground scatterers, all propagation paths can be regarded as nearly parallel, leading to a negligible Doppler spread. Thus, only the common Doppler frequency offset $f_\text{d}$ is considered. 
To simplify \eqref{eq:recvd}, we define the channel frequency response $H(k)$ for subcarrier $k$ as follows:
\begin{equation}
    \label{eq:chan_freq}
    H (k) = \sum_{l=0}^{L-1} h[l] e^{-j \frac{2 \pi k l}{N}}.
\end{equation}
By substituting \eqref{eq:chan_freq} into \eqref{eq:recvd}, $y[n]$ at the receiver before the FFT operation is given by
\begin{equation}
     \label{eq:simpl_y_n}
    y[n] =\frac{1}{\sqrt{N}}  \sum_{k=0}^{N -1} X(k) H(k) e^{j (\frac{2 \pi k  n}{N}+2\pi f_\text{d} T_s n)} + y_{\mathrm{ISI}}[n] + w[n]. \nonumber
\end{equation}
The receiver discards the CP and performs an FFT on the remaining $N$ samples for $n = 0, 1, \ldots, N-1$. After the FFT operation, the received signal at the $m$-th subcarrier is
\begin{align}
\label{eq:after_fft}
    Y(m) = \frac{1}{\sqrt{N}} \sum_{n=0}^{N-1} y[n] e^{-j \frac{2\pi mn}{N}}.
\end{align}
%\begin{comment}
\begin{figure*}[!t]
  \centering
  \subfloat[ICI and the loss in the desired signal power value (blue vs. red at $f_k$)]{%
    \includegraphics[width=0.32\textwidth]{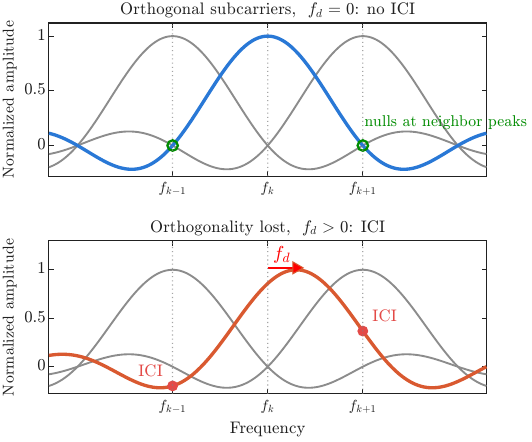}%
    \label{fig:concept_ici}}
  \hfil
  \subfloat[Phase offset]{%
    \includegraphics[width=0.32\textwidth]{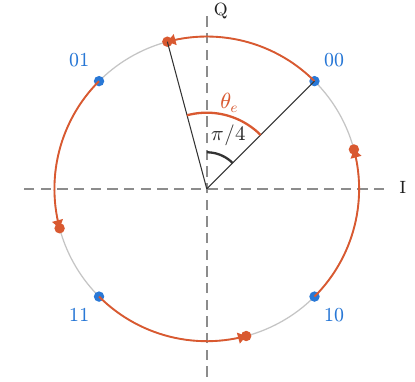}%
    \label{fig:concept_rotation}}
    \hfill
  \subfloat[ISI]{%
    \includegraphics[width=0.32\textwidth]{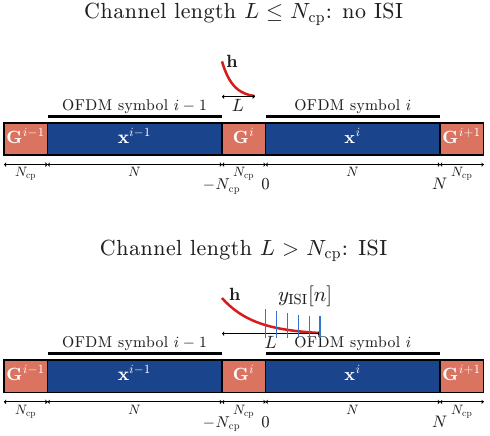}%
    \label{fig:concept_isi}}
\caption{Illustration of the key factors affecting the OFDM signal over a satellite-to-ground sub-THz wireless link besides the atmospheric/spreading losses.}
  \label{fig:concept}
  \vspace{-3mm}
\end{figure*}

\subsubsection*{Inter-carrier Interference}
\label{subsec:ici_derivation}
The Doppler frequency shift $f_\text{d}$ induces the ICI among subcarriers in OFDM systems. As illustrated in Fig.~\ref{fig:concept_ici}, ideally ($f_\text{d} = 0$), each subcarrier has nulls at all other subcarrier frequencies. Meanwhile, any Doppler shift $f_\text{d} \neq 0$ breaks this alignment, as nonzero sidelobe values are sampled at neighboring subcarriers. As consecutive OFDM symbols are statistically independent, the Doppler-induced and multipath-induced ICI components are uncorrelated and analyzed independently below.

Given the bandwidth $B$, the sampling time is $T_s = 1/B$ under Nyquist sampling. Accordingly, \eqref{eq:after_fft} is rewritten as
\begin{align}
    \label{eq:after_fft2}
    Y (m) = \sum_{k=0}^{N-1} X(k) H(k) \left[ \frac{1}{N} \sum_{n=0}^{N-1}
    e^{j \frac{2 \pi (k - m ) n}{N} + j\frac{2\pi f_\text{d} n}{B}} \right]
    \nonumber \\ + \YISI(m) + W(m).
\end{align}

Here, as in \cite{armstrong1999analysis,zhao2001intercarrier}, the term in brackets in \eqref{eq:after_fft2} is defined as the ICI coefficient $C(m,k)$, given by
\begin{align}
    \label{eq:C_m_k}
   C(m,k) &= \frac{1}{N} \sum_{n=0}^{N-1} \left( e^{j \frac{2 \pi (k - m)}{N} + j\frac{2 \pi f_\text{d}}{B}} \right)^n \nonumber \\
          % &= \frac{1}{N} \frac{1 - e^{j 2 \pi (k - m + \epsilon)}}{1 - e^{j \frac{2 \pi (k - m + \epsilon)}{N}}} \nonumber \\
           & =  \frac{\sin(\pi (k - m + \frac{ f_\text{d} N}{B}))}{N \sin(\frac{\pi (k - m + \frac{ f_\text{d} N}{B})}{N})} e^{j \pi (k - m + \frac{  f_\text{d} N}{B}) \left( 1 - \frac{1}{N} \right)}
\end{align}

By explicitly separating the desired signal and ICI impact, the received signal in \eqref{eq:after_fft2} can be reformulated as
{\allowdisplaybreaks
\begin{align}
    %\label{eq:ICI_description}
    &Y(m) = \YS(m) + \YICI (m)+ \YISI (m) + W(m) \nonumber \\ 
    & \YS(m) = X(m) H(m) C (m,m) \label{eq:desired_sig}\\
    &\YICI (m) = \sum_{k=0, k \neq m}^{N-1} X(k) H(k) C(m,k),
    \label{eq:ICI_sig}
\end{align}}
where $\YS (m)$ and $\YICI (m) $ denote the desired and Doppler-induced ICI signals for subcarrier $m$, respectively.

%%%%%%%%%%
\subsubsection*{Desired Signal Power} 
\label{subsec:desired_signal_power_analysis}
The presence of ICI in non-zero Doppler scenarios means that a part of the received energy from the selected subcarrier becomes interference at other subcarriers. Hence, the power of the desired signal itself over a given subcarrier reduces accordingly.
From \eqref{eq:desired_sig} and following the approach in~\cite{pollet1995ber}, the average desired signal power over the channel $\boldsymbol{h}$ is approximated as follows:
\begin{align}
     \label{eq:Ps_approx}
    \bar{\Ps}  
 \approx 1- \frac{1}{3}\left(\frac{\pi N f_\text{d}}{B}\right)^2,
\end{align}
where, for clarity, only the small-scale fading $\boldsymbol{h}$ is considered (i.e., the large-scale path loss factor $1/\sqrt{P_{\rm L}}$ in the channel model of Section~\ref{subsec:channel_model} is, for now, factored out).

%%%%%%%%%%
\subsubsection*{ICI Power}
\label{subsec:ici_power}
 From \eqref{eq:ICI_sig}, the average ICI $\PIDP$ induced by the Doppler shift is given by
\begin{align}
\label{eq:P_ICI_approx}
  \PIDP \triangleq 1- \bar{P}_\text{S} \approx  \frac{1}{3 }\left(\frac{\pi N f_\text{d}}{B}\right)^2.
\end{align}

%%%%%%%%%%
\subsubsection*{Phase offset}
\label{subsec:phase_rot}
On top of the ICI impairments, the presence of a non-zero Doppler frequency shift induces a deterministic phase offset, which can cause phase synchronization and demodulation errors~\cite{kim2007residual}. To illustrate, Fig.~\ref{fig:concept_rotation} shows the effect of the phase offset on the QPSK constellation. All QPSK symbols rotate rigidly by the phase offset $\theta_e$, and when $\theta_e$ exceeds the phase margin of $\pi/4$, the rotated symbols cross the decision boundaries, causing symbol errors. Hence, the phase offset must be carefully accounted for in the design of OFDM systems for sub-THz LEO satellite links.

From~\eqref{eq:C_m_k}, when $m = k$, the phase offset of the desired signal is given by $\pi f_\text{d} (N-1)/B$. Moreover, given~\eqref{eq:recvd} and the OFDM symbol length of $N+\Ncp$ samples, the phase offset is accumulated by  $2\pi f_\text{d} (N+ \Ncp)/B$ per OFDM symbol. Thus, the cumulative phase offset after $q$ OFDM symbols, including the phase offset from the desired signal, is expressed as
\begin{equation}
\label {eq:phase_rotation_accum}
\theta_e (q) = \frac{\pi f_\text{d} (q) (N-1)}{B} + \sum_{i=1}^q \frac{2\pi f_\text{d} (i)(N+\Ncp)}{B},
\end{equation}
where $f_\text{d}(i)$ denotes the Doppler shift experienced at OFDM symbol $i$. Equation \eqref{eq:phase_rotation_accum} indicates that a large $N$ increases the phase offset, particularly for large Doppler shifts.
%%%%%%%%%%%%%%%%%%%%%%%%%%%%%%
\subsection{ISI Analysis}
\label{subsec:isi_analysis}
The signal passing through a modeled sub-THz satellite communication link is affected not only by Doppler-induced effects, but also by other major factors. These particularly include the spreading loss, accounted for in the channel model detailed in Section~\ref{sec:system_model}, and the multipath-induced ISI effects illustrated in Fig.~\ref{fig:concept_isi} and discussed below.

In the modeled scenario, the ISI is caused by the preceding OFDM symbol when the length $\Ncp$ of the CP is less than that of the multipath channel impulse response $L$. Hence, the $(i-1)$-th OFDM symbol interferes with the $i$-th symbol when $\Ncp < L $. For a channel length $L \le N_{\mathrm{cp}}$, the tail of symbol $i{-}1$ is absorbed within the CP, whereas for $L > N_{\mathrm{cp}}$ its residual intrudes into the FFT window of symbol~$i$. The resulting ISI term is given by
\begin{align}
y_{\mathrm{ISI}}[n] &= \sum_{\nu = N_{\mathrm{cp}}+1+n}^{L-1}h[\nu]  x^{i-1}[n-\nu]  \\
&\qquad 0 \le n \le L - N_{\mathrm{cp}} - 2 \nonumber
\end{align}

For notational brevity, the OFDM symbol index $(i-1)$ is dropped in the following derivations.
Based on the approach in~\cite{HenkelCyclicPrefix2002}, the ISI at the $m$-th subcarrier becomes
\begin{align}
&\YISI(m)  = \frac{1}{\sqrt{N}}\sum_{p=\Ncp+1}^{L-1}x[-p]e^{j\frac{2\pi}{N}pm}H_{p}(m),
\end{align}
where $H_{p}(m) = \sum_{\mu=p}^{L-1}h[\mu]e^{-j\frac{2\pi}{N}\mu m} $.

The power of ISI corresponding to the $m$-th subcarrier is
\begin{align}
%\label{eq:P_ISI}
\PISI(m)
= \mathbb{E} \left[ \YISI(m) \YISI^*(m) \right]
= \frac{1}{N}
\sum_{p = \Ncp+1}^{L -1}
\left| H_p(m) \right|^2. \nonumber
\end{align}
In addition, multipath propagation gives rise to ICI by destroying the orthogonality among subcarriers. Following~\cite{HenkelCyclicPrefix2002, batariere2004cyclic}, the multipath-induced ICI power is approximately equal to the ISI power when $0 < L - N_{\text{cp}} \ll N$, so that the total interference power due to multipath propagation at subcarrier $m$ is approximately $2P_{\text{ISI}}(m)$.

It is worth noting that $P_{\text{ISI}}(m)$ significantly exceeds the ICI power due to multipath effects when $L - N_{\text{cp}} \gg N$~\cite{batariere2004cyclic}. Therefore, $2P_{\text{ISI}}(m)$ represents the worst-case \emph{multipath-induced} interference power in general, and its average, denoted as $\PIMP \triangleq \mathbb{E}_{\boldsymbol{h}} [2\PISI] $, if only the normalized small-scale channel is considered, becomes
{\allowdisplaybreaks
\begin{align}
\label{eq:P_ISI_avg}
\PIMP &=  \frac{2}{N} \sum_{p = \Ncp+1}^{L-1}
\mathbb{E}_{\boldsymbol{h}}\left[\left| \sum_{\mu=p}^{L-1} h[\mu] e^{-j \frac{2\pi}{N} \mu m} \right|^2 \right] \nonumber \\
&= \frac{2}{N}  \sum_{p = \Ncp+1}^{L-1}\sum_{\mu=p}^{L-1} \mathbb{E}_{\boldsymbol{h}}[|h[\mu]|^2] \nonumber \\
&=\frac{2}{N}  \sum_{p = \Ncp+1}^{L-1}\frac{(L - p)}{L(K+1)} 
\approx \frac{1}{N}   \frac{(L - \Ncp)^2}{L(K+1)}.
\end{align}}
%%%%%%%%%%%%%%%%%%%%%%%%%%%%%%%%%%%%%%%%%%%%%%%%%%
\subsection{Key Performance Indicators}
\label{subsec:kpi_analysis}
We first derive closed-form expressions for the average SINR. We then formulate effective throughput, which serves as the key performance metric (KPI) for evaluating different subcarrier spacing (SCS) configurations in Section~\ref{sec:numerical_results}.

%%%%%%%%%%
\subsubsection*{Average SINR}
Combining all the individual small-scale effects discussed above with the large-scale path loss $P_{\rm L}$ and applying the Delta Method approximation~\cite{casella2024statistical}, the average SINR $\bar{\gamma}_{\text{SINR}}$ of a subcarrier $m$ becomes
\begin{align}
\bar{\gamma}_{\mathrm{SINR}} (m) %&=
&\approx \frac{P_m \bar{\Ps} P_{\rm L}^{-1}}{P_m  (\PIDP + \PIMP)P_{\rm L}^{-1} + N_0 \Delta f N_f} \label{eq:SINR_approx},
\end{align}
where $N_f$ is noise figure, $N_0$ is noise power spectral density, $\Delta f$ is subcarrier spacing, given as $B/N$, $P_m$ denotes the power allocated to subcarrier $m$, and $P_\text{L}$ is free space path loss, as given in \eqref{eq:pathloss}. 
%In ~\eqref{eq:SINR_approx}, the large-scale path loss $P_{\rm L}$ -- factored out of the normalized small-scale channel used to derive $\bar{\Ps}$, $\PIDP$, and $\PIMP$ in (14), (15), and (19) -- is reintroduced here through the multiplicative factor $P_{\rm }^{-1}$.
%, so that \eqref{eq:SINR_approx} jointly captures the small-scale Doppler- and multipath-induced impairments together with the large-scale propagation loss.
The approximation in \eqref{eq:SINR_approx} can be justified in LOS-dominant sub-THz LEO channels due to their relatively limited interference power fluctuation.

Finally, by substituting \eqref{eq:P_ISI_avg}, \eqref{eq:Ps_approx}, \eqref{eq:P_ICI_approx}, and \eqref{eq:pathloss} into \eqref{eq:SINR_approx}, the closed-form expression for the average SINR $\bar{\gamma}_{\text{SINR}}$ in our modeled scenario becomes
\begin{equation}
\begin{aligned}
\bar{\gamma}_{\text{SINR}} =
\frac{1 - \frac{1}{3}\left(\frac{\pi N f_\text{d}}{B}\right)^2}
{\frac{1}{3}\left(\frac{\pi N f_\text{d}}{B}\right)^2 + \frac{(L-N_{\text{cp}})^2}{NL(K+1)} + \frac{N_0 N_f B L_\text{A}}{P_\text{T} G_\text{T} G_\text{R}}\left(\frac{4\pi\rho f_\text{c}}{c}\right)^2} 
\end{aligned}
\end{equation}

%%%%%%%%%%
\subsubsection*{Average Effective Throughput}
For clarity, we consider the average effective throughput over the random channel $\boldsymbol{h}$ without any channel coding.
The instantaneous effective throughput for a given channel realization $\boldsymbol{h}$ is given by~\cite{proakis2001digital}
\begin{equation}
\label{eq:R_eff}
R_{\text{eff}}(\boldsymbol{h}) = \left(1 - P_{\text{block}}(\boldsymbol{h})\right) \frac{NB\log_2 M}{N + N_{\text{cp}}}.
\end{equation}
Averaging over $\boldsymbol{h}$ yields the average effective throughput
\begin{equation}
\label{eq:R_eff_avg}
\bar{R}_{\text{eff}} = \left(1 - \bar{P}_{\text{block}}\right) \frac{NB\log_2 M}{N + N_{\text{cp}}},
\end{equation}
where $\bar{P}_{\text{block}} = \mathbb{E}_{\boldsymbol{h}}\left[P_{\text{block}}(\boldsymbol{h})\right]$ is the average block error rate (BLER).

\subsubsection*{Average Error Rate} 
%Assuming that all subcarriers experience identical channel conditions within one transmission block, a uniform bit error rate (BER) $P_b^M(\boldsymbol{h})$ is applied across all subcarriers. 
\textcolor{black}{Under the assumption of the strong LOS-dominant propagation, all subcarriers are taken to experience similar channel conditions within one transmission block, so that a uniform bit error rate (BER) $P_b^M(\boldsymbol{h})$ is applied across all subcarriers.}
Accordingly, the BLER $P_{\text{block}}$ is $P_{\text{block}}(\boldsymbol{h}) = 1 - \left(1 - P_b^M (\boldsymbol{h})\right)^Q$, where $Q$ denotes the block size of \emph{uncoded} data.

When employing M-ary phase-shift keying (M-PSK) modulation with order $M$ and phase offset $\theta_e$, the closed form of the symbol error rate (SER) $P_{\text{sym}}$ becomes~\cite{simon1998some}
\begin{equation}
\begin{aligned}
P_\text{sym}^{M} (\boldsymbol{h}) &= \frac{1}{2\pi}
\int_{0}^{\frac{(M-1)\pi}{M}-\theta_e}
\exp\left(
-\gamma_{\text{SINR}}
\frac{\sin^2\left(\frac{\pi}{M}+\theta_e \right)}
{\sin^2\Phi}
\right) d\Phi
\\ %\nonumber
&+
\frac{1}{2\pi}
\int_{0}^{\frac{(M-1)\pi}{M}+\theta_e}
\exp\left(
-\gamma_{\text{SINR}}
\frac{\sin^2\left(\frac{\pi}{M}-\theta_e\right)}
{\sin^2\Phi}
\right) d\Phi,
\end{aligned}
\end{equation}
where $\theta_e$ is given in~\eqref{eq:phase_rotation_accum} and $\gamma_{\text{SINR}}$ is the instantaneous SINR. Accordingly, in the case of Gray mapping, the average BER is approximated by $P_b^{M}(\boldsymbol{h}) \approx P_\text{sym}^{M}(\boldsymbol{h})/\log_2 M$.

Importantly, from the analysis above, a greater number of OFDM subcarriers $N$ has both positive (lower $\PIMP$) and negative (higher $\PIDP$ and lower $\bar{\Ps}$) effects on the KPIs. Therefore, selecting an appropriate number of subcarriers $N$ for the given scenario configuration becomes one of the major design considerations, as discussed in the next section.

%%%%%%%%%%%%%%%%%%%%%%%%%%%%%%%%%%%%%%%%%%%%%%%%%%%%%%%%%%%%%%%%%%%%%%%%%%%%%%%%%%%%%%%%%%%%%%%%%%%%%%%%%%%%%
\section{Numerical Results}
\label{sec:numerical_results}

\begin{table}[!b]
\centering
\caption{Main Numerical Parameters}
\label{tab:simulation_parameters}
\renewcommand{\arraystretch}{1.2}
\begin{tabular}{l|c|c}
\hline
\textbf{Parameter} & \textbf{Notation} & \textbf{Default value} \\
\hline
\hline
Carrier frequency & $f_\text{c}$ & \SI{220}{GHz} \\
\hline
System bandwidth & $B$ &  \SI{5}{GHz} \\
\hline
Satellite altitude & $h_\text{S}$ & \SI{600}{km} \\
\hline
Transmitted power  & $P_\text{T}$ & \SI{27}{dBm}~\cite{cooper2025power}\\
\hline
Satellite max Tx antenna gain & $G_\text{T}$ & \SI{46.8} {dBi} \\
\hline
Ground user max Rx antenna gain & $G_\text{R}$ & \SI{63.6} {dBi}\\
\hline
Noise spectral density & $N_0$ & \SI{-174}{dBm/Hz} \\
\hline
Noise figure  & $N_f$ & \SI{7}{\decibel} \\
\hline
Modulation scheme & -- & BPSK \\
\hline
Block size & Q & \SI{24}{bit}\\
\hline
Orbit inclination angle & $i_{\rm o}$ & $53^\circ$ \\
\hline
\end{tabular}
\end{table}

\begin{figure*}[!t]
    \centering   
    \subfloat [Doppler-induced, $\PIDP$, and multipath-induced, $\PIMP$, interference]{
        \includegraphics[width=0.95\columnwidth]{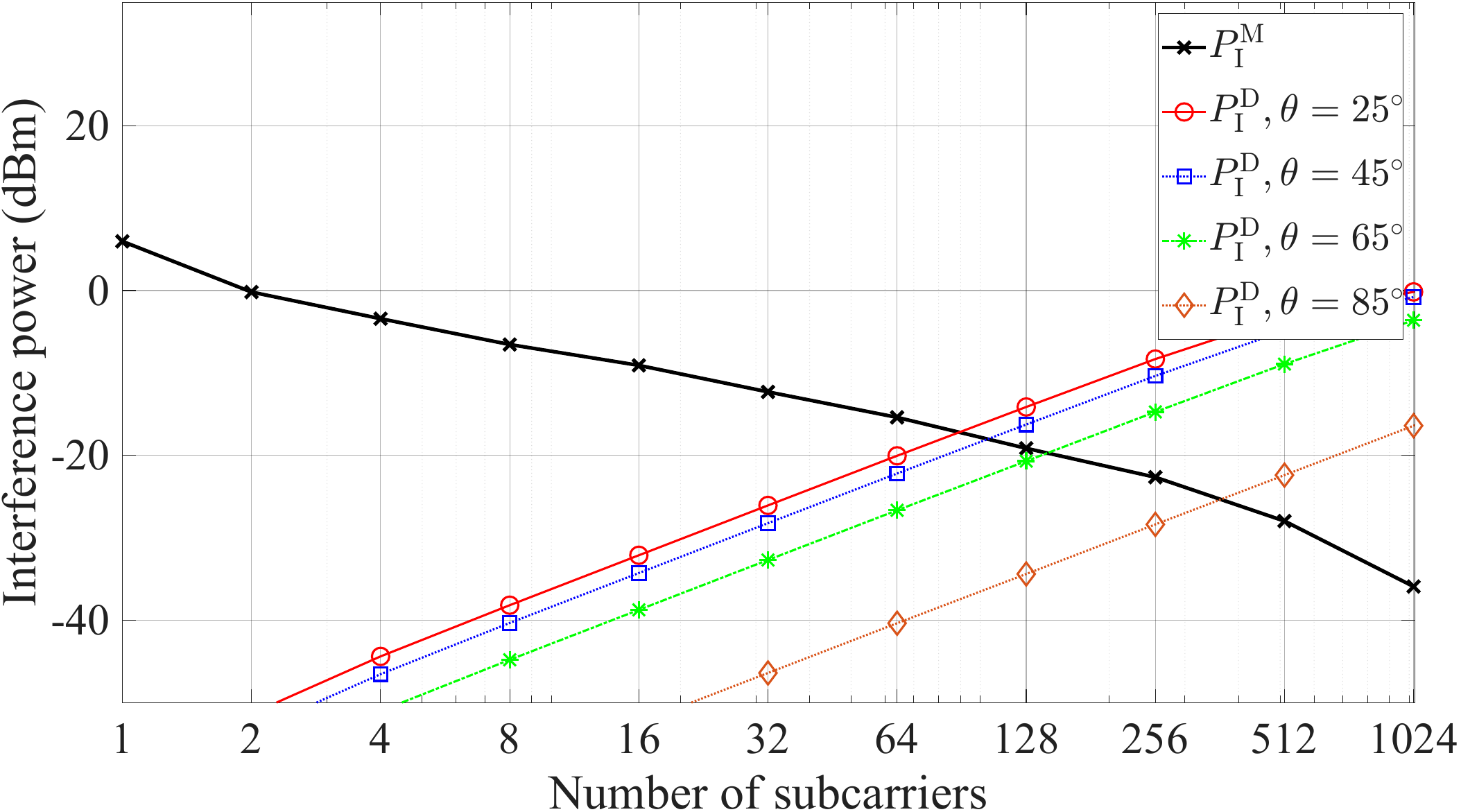}
        \label{fig:interf_power_different_doppers}
        }
    \subfloat[Average SINR]{
        \includegraphics[width=0.95\columnwidth]{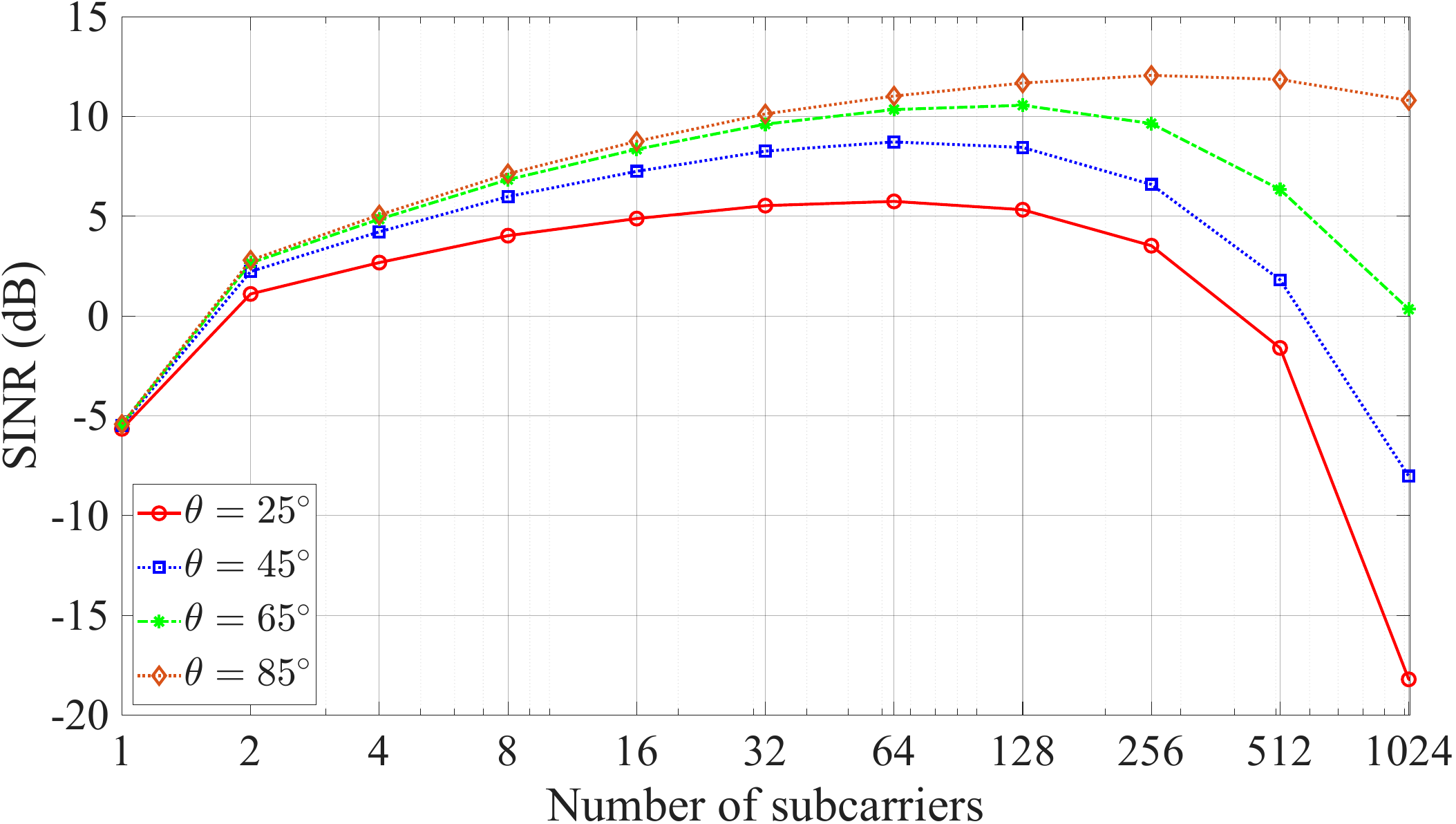}
        \label{fig:SINR_different_doppers}
        }
    \caption{Average interference power and SINR for different values of elevation angles $\theta$ and number of subcarriers $N$.}
    \label{fig:P_ICI_SIR_diff_Dopplers}
    \vspace{-5mm}
\end{figure*}

%are then computed based on~\cite{maral2020satellite}

Our numerical study is conducted over the range of satellite elevation angles from $\theta_{\min} = 25^{\circ}$ to $\theta_{\max} = 90^{\circ}$~\cite{fcc2022grantSpaceX}. The channel is modeled as an LOS-dominant channel with a Rician K-factor of $\SI{17}{\dB}$ and a delay spread of $\SI{10}{\nano\second}$~\cite{ITU_M2541_2024}. Under sampling time $T_s = 1/B$, a CP overhead of approximately $7\%$, i.e., $N_{\text{cp}} = \lfloor 0.07N \rfloor$, and a block size $Q$ is set according to~\cite{3gpp38214}. 
\textcolor{black}{The satellite ($\SI{10}{\centi\meter}$, CubeSat-typical) and GU ($\SI{70}{\centi\meter}$) antenna diameters yield the directive gains in Table~\ref{tab:simulation_parameters}, which are computed based on ~\cite{maral2020satellite} by assuming an aperture efficiency of $0.9$.} In the numerical study, we deliberately set $L_\text{A} = 1$ to exclude the additional time-varying propagation losses (such as atmospheric absorption~\cite{Aliaga2025Scintillation} and rain attenuation~\cite{ma2018effect,sen2022terahertz}), while focusing primarily on the impact of Doppler shift on the satellite-to-ground sub-THz link with sufficient antenna gains to close the link budget. Remaining parameters are summarized in Table~\ref{tab:simulation_parameters}.
%\textcolor{red}{[Reviewer 2] 4.	Table I The assumption to use just BPSK modulation simplifies the analysis but it is far from reality because the most recent technological initiatives (e.g. Starlink) adopts high level modulation to get higher bit rate.}
%%%%%%%%%%%%%%%%%%%%%%%%%%%%%%%%%%%%%%%%%%%%%%%%%%
\subsection{Impact of the Number of Subcarriers and Elevation Angle}
We start with Fig.~\ref{fig:P_ICI_SIR_diff_Dopplers} exploring the average interference power (Doppler-induced and non-Doppler-induced) and average SINR for different values of elevation angles $\theta$ and the number of subcarriers $N$. Specifically, Fig.~\ref{fig:interf_power_different_doppers} presents the average interference powers induced by the Doppler shift and by multipath propagation, denoted by $\PIDP$ and $\PIMP$, respectively, as a function of the number of subcarriers $N$.

% [!!!]Question 1 to SJ: Why does $\PIMP$ in Fig. 3(a) in this setup depend \emph{only} on the number of subcarriers $N$ but not on the elevation angle $\theta$? The equation depends on the path loss, which is different for different elevation angles, no? Why is the selected range of 25, 45, 60 (not 65!!!), 85 selected in such a manner?? Where is 90 degrees, for instance?

In this figure, as $N$ increases, $\PIDP$ grows while $\PIMP$ diminishes, as expected from \eqref{eq:P_ICI_approx} and \eqref{eq:P_ISI_avg}. Moreover, since Doppler shift increases as the elevation angle decreases, as indicated by \eqref{eq:doppler_gamma}, the higher value of $\PIDP$  is observed at lower elevation angles.
Correspondingly, the average SINR $\bar{\gamma}_{\text{SINR}}$ in similar conditions presented in Fig.~\ref{fig:SINR_different_doppers} initially increases with the number of subcarriers because $\PIMP$ decreases when $N$ increases, as indicated by \eqref{eq:P_ISI_avg}. When $\PIDP$ becomes nearly equal to $\PIMP$ in Fig.~\ref{fig:interf_power_different_doppers}, e.g., for $N = 64$ or $N = 128$ setups, the average SINR remains above $\SI{5}{\decibel}$ across all elevation angles---the highest SINR among all the considered values of $N$. However, when $N$ gets larger, i.e., $N \geq 256$, the SINR degrades substantially at lower elevation angles.

\begin{figure*}[!t]
    \centering   
    \subfloat [Average SINR]{
        \includegraphics[width=0.99\columnwidth]{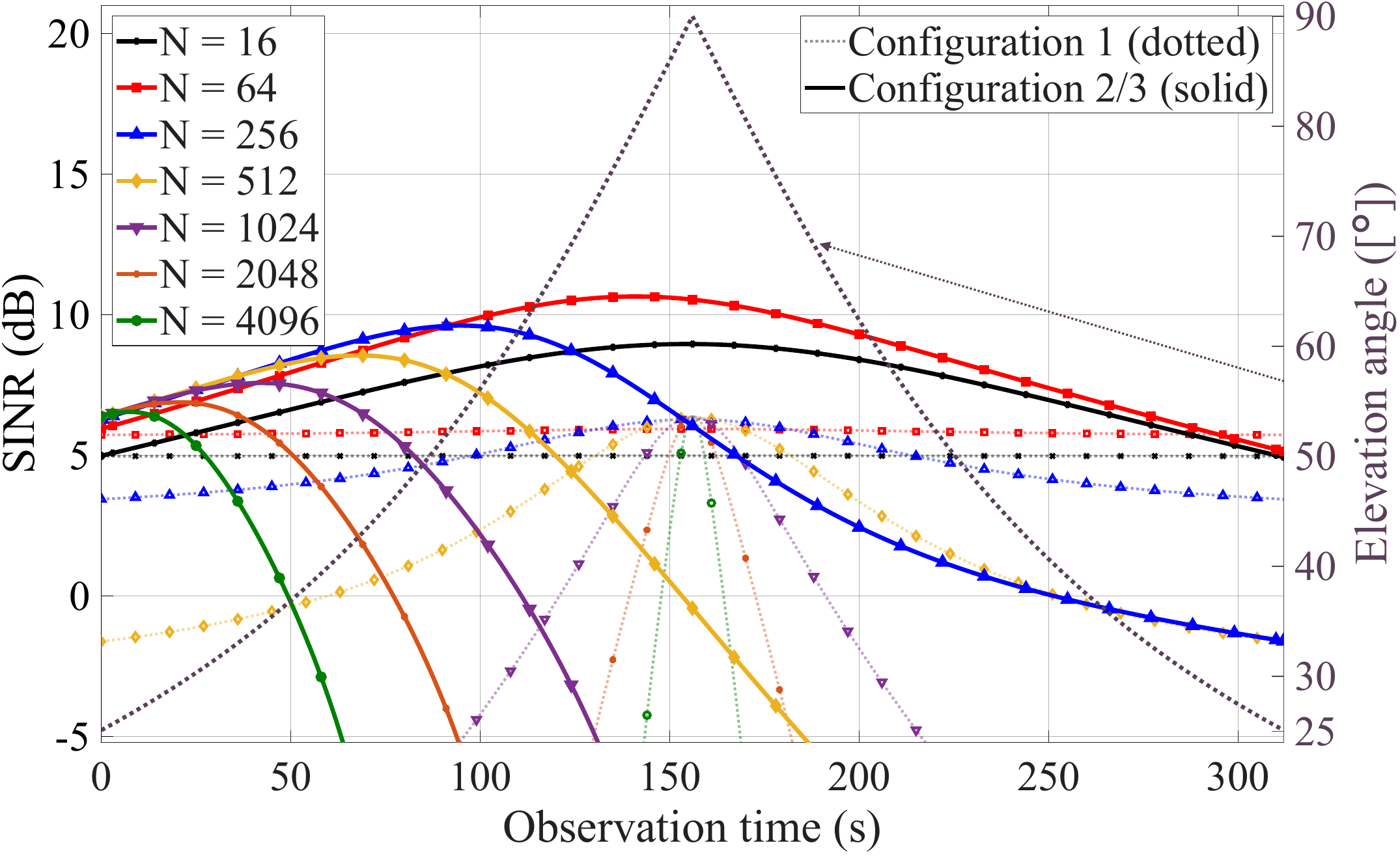}
        \label{fig:SINR_vs_time}
        }
    \subfloat[Effective throughput]{
        \includegraphics[width=0.99\columnwidth]{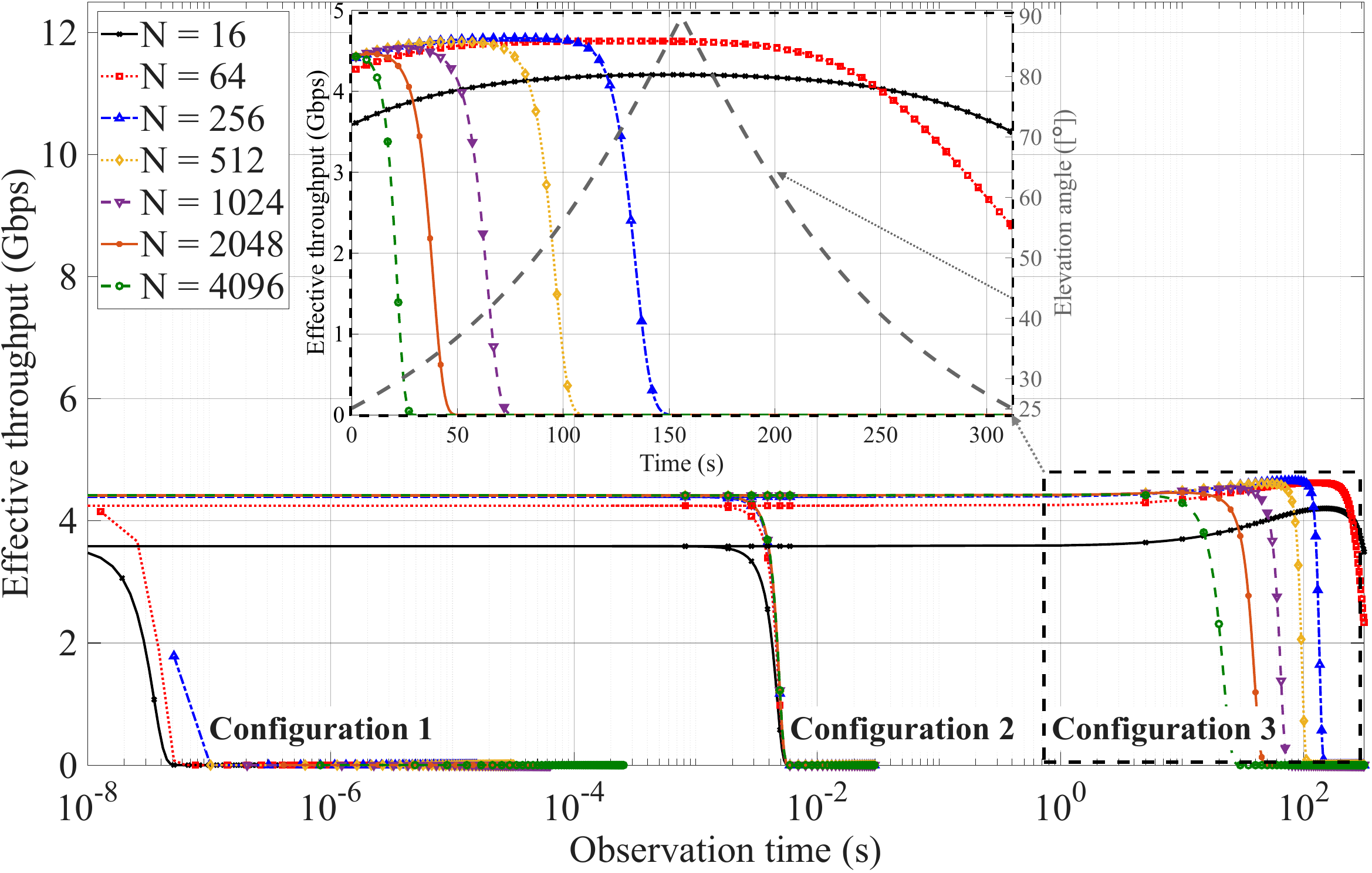}
        \label{fig:eff_tput_bpsk_5}
        }
    \caption{Average SINR and effective throughput for different Doppler and phase pre-compensation configurations (summarized in Table~\ref{tab:sys_confg}).}
    \label{fig:Doppler_time}
    \vspace{-3mm}
\end{figure*}

\begin{comment}
\begin{figure*}[!t]
    \centering   
    \subfloat [Average SINR]{
        \includegraphics[width=0.95\columnwidth]{figures/SINR_vs_theta_each_config_time_BPSK.pdf}
        \label{fig:SINR_vs_time}
        }
    \subfloat[Effective throughput]{
        \includegraphics[width=0.95\columnwidth]{figures/effTput_vs_theta_each_config_time_BPSK.pdf}
        \label{fig:eff_tput_bpsk_5}
        }
    \caption{Average SINR and effective throughput for different Doppler and phase pre-compensation configurations (summarized in Table~\ref{tab:sys_confg}).}
    \label{fig:Doppler_time}
    \vspace{-3mm}
\end{figure*}
\end{comment}

\begin{table}[!b]
\vspace{-5mm}
\centering
\caption{Compared Doppler-agnostic and Doppler-aware configurations\vspace{-2mm}}
\label{tab:sys_confg}
\begin{tabular}{c|c|c}
\hline
Configuration & Frequency pre-compensation & Phase pre-compensation \\
\hline
\hline
$1$ & -- & -- \\
\hline
$2$ & \checkmark & -- \\
\hline
$3$ & \checkmark & \checkmark \\
\hline
\end{tabular}
\end{table}
%%%%%%%%%%%%%%%%%%%%%%%%%%%%%%%%%%%%%%%%%%%%%%%%%%
\subsection{Impact of Doppler Compensation Approaches}
Next, the effects of Doppler shift on the time-varying effective throughput, as defined in~\eqref{eq:R_eff_avg}, are examined in Fig.~\ref{fig:Doppler_time} over a prolonged observation time interval, as the LEO satellite travels along its orbit over the GU. For the modeled setup with $h_\text{S} = \SI{600}{\kilo \meter}$, the path takes approximately $\SI{304}{\second}$, during which the elevation angle $\theta$ varies from $\theta_{\min} = 25^\circ$ to $\theta_{\max} = 90^\circ$ and back to $\theta_{\min}$.
%the path takes approximately $\SI{315}{\second}$, corresponding to the elevation angle $\theta$ variations between $\theta_{\min} = 25^{\circ}$ and $\theta_{\max} = 90^{\circ}$.

To compare different possible strategies of dealing with Doppler-related effects in such a system (from ignoring to accounting for), three system configurations are considered, as summarized in Table~\ref{tab:sys_confg}. 
Particularly, in Configuration~$1$, no Doppler frequency or phase pre-compensation is applied at the transmitter/receiver. In Configurations~$2$ and $3$, the Doppler frequency estimated at the initial transmission time ($\theta = \theta_{\text{min}}$) is used for pre-compensation \emph{throughout the observation interval}. \textcolor{black}{The \emph{residual} Doppler shift $\Delta f_{\mathrm d} (t)$ at time $t$ for Configuration 2 and 3 is represented as 
\begin{equation}
    \Delta f_{\mathrm d} (t)  =
    f_{\mathrm d}(t) - \hat{f_{\mathrm d}},
    \label{eq:residual_Doppler}
\end{equation}
where $\hat{f_{\mathrm d}}$ denotes the Doppler shift perfectly estimated at $t=0$, i.e., $\hat{f_{\mathrm d}} \triangleq f_{\mathrm d} (t=0)$, which can be obtained, for example, using the method in~\cite{3GPP38821}.}
\textcolor{black}{For Configuration 2, since there is no phase pre-compensation, the phase offset is accumulated with the residual Doppler shift $\Delta f_\text{d}$ according to~\eqref{eq:phase_rotation_accum}.}
In contrast to Configuration~2, Configuration~3 also applies linear phase pre-compensation on a per-symbol basis to balance the accumulating phase rotation (and corresponding decoding errors discussed in Section~\ref{subsec:doppler_effect_analysis}).
\textcolor{black}{In particular, for Configuration $3$ after linear phase pre-compensation, the residual phase offset $\Delta \theta_e$ of the $q$-th symbol at time $t$ is represented by the residual Doppler shift $\Delta f_{\mathrm d}(t)$ as
\begin{equation}
\Delta \theta_e (q)=\theta_e(q) - \theta_e (q-1) = \frac{2\pi \Delta f_{\mathrm d} (t) (N+\Ncp)}{B} 
%+  \frac{2\pi \Delta f_\text{d} (i)N}{B}    
\label{eq:phase_rotation_accum_Conig3}
\end{equation}}
Notably, to facilitate a fair comparison, all the remaining parameters stay the same for all three configurations. For Configurations 2 and 3, the residual Doppler shift in~\eqref{eq:residual_Doppler} increases over time.

%\textcolor{red}{[Reviewer 1] The authors shall explain the algorithm used for Doppler estimation and indicate the residual error at t=0 when compensation is applied.}
\textcolor{black}{
Proceeding with Fig.~\ref{fig:SINR_vs_time}, Configurations~2 and~3 (solid lines) share an identical
SINR at $t = 0$ (i.e., $\theta = \theta_{\text{min}}$), since both apply the same initial frequency pre-compensation.
Configuration~1 (dotted line), which applies no compensation at all, instead tracks the instantaneous Doppler shift $f_{\mathrm d}(t)$ directly. Therefore, its SINR is lowest near the pass edges, where the elevation angle approaches $\theta_{\text{min}}$ and $|f_{\mathrm d}(t)|$ is largest, but recovers considerably near the zenith crossing ($\theta \to 90^{\circ}$, $t \approx 152$~s), where $f_{\mathrm d}(t) \to 0$.
For a large number of subcarriers $N$, Configuration~1 can even temporarily exceed Configurations~2 and~3 near the zenith crossing, as the residual Doppler shift in Configurations~2 and~3 grows consistently over time.}
%Overall, however, Configuration~3 maintains the most favorable SINR throughout the majority of the observation window due to its continuous phase-tracking compensation.

%\textcolor{black}{Proceeding with Fig.~\ref{fig:SINR_vs_time}, the SINR in Configuration~$1$ is observed to be the lowest among all configurations, caused by a large Doppler shift and the resulting strong Doppler-imposed ICI, particularly for a large number of subcarriers $N$.
%In contrast, Configurations~$2$ and Configuration~$3$ (both featuring Doppler shift estimation at $t = 0$\, or $\theta = \theta_{\text{min}}$) exhibit the same and much more favorable SINR values over time since they experience the same residual Doppler shift as in~\eqref{eq:residual_Doppler}.}  
%\textcolor{black}{On the other hand, Fig.~\ref{fig:SINR_vs_time} shows the SINR variation over time until the effective throughput drops due to accumulated phase rotation, as will be further discussed in Fig.~\ref{fig:eff_tput_bpsk_5}. Notably, no such SINR degradation is observed for Configuration~1 within the considered time interval.}
%\textcolor{red}{[Reviewer 1] Why in Fig.4a, the SINR degradation for config1 and config2 are not shown? }
%\textcolor{red}{ [Reviewer 1] a brief explanation on how residual doppler error will impact the SINR could be added to better interpret the results in Fig. 4a. }

At the same time, an essential difference between the last two configurations arises when studying the effective throughput in Fig.~\ref{fig:eff_tput_bpsk_5}. Here, with Configuration~$3$, the effective throughput increases over time, due to the linear pre-compensation of the phase offset (hence, BER is growing much slower over time) and a general reduction of the communication distance from $t = 0$\,s to $t = 152$\,s, as the satellite travels over the orbit between $\theta_\text{min}$ and $\theta_{\text{max}}$ elevation angles. For the scenario geometry in Fig.~\ref{fig:Earth_and_groundUser}, the communication paths closer to zenith are more favorable for the data exchange featuring shorter $\rho$ and thus lower $P_{\text{L}}$.

Nevertheless, even for the most advanced Configuration~$3$, the SINR and effective throughput still degrade significantly for a larger number of subcarriers $N$ since the increase of the residual Doppler shift results in strong ICI and the increase of phase offset over time. Toward the middle of the observation interval at approximately $t = \SI{152}{\second}$ (note that the time axis in Fig.~\ref{fig:eff_tput_bpsk_5} is in log scale), Configuration~$3$ exhibits a throughput increase for a smaller number of subcarriers, i.e., when $N \leq 64$, since the received SNR increases as the satellite approaches $\theta_{\max}$.
%\textcolor{black} {After the satellite passes the maximum elevation angle (after $\SI{155}{\second}$), it is observed that the effective throughput decreases since the SNR starts decreasing with the increase in distance and the increase in residual Doppler shift $\Delta f_\text{d}$ over time.}
\textcolor{black}{After the satellite reaches its maximum elevation angle (i.e., for $t > \SI{152}{\second}$), the effective throughput decreases. This degradation is attributed to two factors: the SNR decreases as the distance increases, and the residual Doppler shift $\Delta f_\text{d}$ grows, which intensifies the ICI.}

%the sign of the Doppler shift $f_\text{d}$ becomes opposite, causing the accumulated phase offset to decrease.}
%\textcolor{black} {After the satellite passes the maximum elevation angle (after $\SI{155}{\second}$), the sign of the Doppler shift $f_\text{d}$ becomes opposite, causing the accumulated phase offset to decrease.}

%%%%%%%%%%%%%%%%%%%%%%%%%%%%%%%%%%%%%%%%%%%%%%%%%%
\subsection{Severity of Doppler Effects in Sub-THz LEO Satellite Links}
Analyzing Fig.~\ref{fig:P_ICI_SIR_diff_Dopplers} and Fig.~\ref{fig:Doppler_time} together from a system design perspective, the following major observations can be derived. First, totally ignoring the Doppler shift in wideband sub-THz LEO satellite links, i.e., Configuration~$1$,
%(by e.g., applying Configuration~1 or selecting an inappropriate number of subcarriers $N$}) 
leads to unacceptably poor performance and reliability. %Specifically, after just a few OFDM symbols transmitted (see e.g., Fig.~\ref{fig:eff_tput_bpsk_5}), the demodulation error rate for Configuration~$1$ grows drastically due to the accumulated phase offset, resulting in close-to-zero effective throughput.
Specifically, for Configuration~1, the demodulation error rate increases sharply after only a few OFDM symbols owing to the accumulated phase offset, which results in a near-zero effective throughput (see, e.g., Fig.~\ref{fig:eff_tput_bpsk_5}).

Second, only estimating the Doppler shift once at the beginning of the communication period at $\theta = \theta_{\min}$ (e.g., within a periodic channel estimation procedure) but simultaneously ignoring the accumulating phase offset (Configuration~2) is not practical either for sub-THz satellite-to-GU wideband wireless links. 
%On paper, SINR curves for Configuration~$2$ are substantially improved compared to Configuration~$1$, 
Compared with Configuration~1, Configuration~2 achieves a substantially higher SINR, and its stable-throughput period in Fig.~\ref{fig:eff_tput_bpsk_5} is up to \emph{as much as five orders of magnitude longer}. 
Meanwhile, the stable-throughput period is only approximately \SI{10}{\milli\second}, which implies that, in practice, the Doppler shift must be re-estimated \emph{every few tens of milliseconds}. 
This challenges the applicability of Configuration~$2$-based setups to practical sub-THz satellite communication systems.

At the same time, estimating the Doppler shift once at the beginning of the communication period at $\theta = \theta_{\min}$ and applying a low-complexity linear phase offset compensation mechanism (as in Configuration~3) leads to substantially better performance. Specifically, with a proper selection of the number of subcarriers $N$ (e.g., $N = 64$ for the particular scenario in Table~\ref{tab:simulation_parameters} and see Fig.~\ref{fig:eff_tput_bpsk_5}), \emph{just a single channel estimation} at $\theta = \theta_{\min}$ becomes sufficient to guarantee robust and high-performance system operation \emph{over the entire duration of the satellite pass over the GU} (approximately $304$ seconds). As a result, in practical deployments, the periodicity of channel estimation and adaptation mechanisms for Configuration~$3$-based setups is to be primarily determined by \emph{non-Doppler-induced factors}, such as atmospheric effects and dynamic beam realignment.

%%%%%%%%%%%%%%%%%%%%%%%%%%%%%%%%%%%%%%%%%%%%%%%%%%%%%%%%%%%%%%%%%%%%%%%%%%%%%%%%%%%%%%%%%%%%%%%%%%%%%%%%%%%%%
\section{Conclusion}
\label{sec:conclusion}
In this work, we quantified the Doppler-shift-related effects on OFDM-based sub-THz LEO satellite-to-ground communications. After reviewing the theory for major Doppler-related signal impairments in such a system, including the Doppler-induced power loss in a given subcarrier, the resulting ICI, and the phase offset, the key performance indicators were derived by accounting for a superposition of these effects for a sub-THz wireless link between a LEO satellite and a ground user. The main conclusions from our numerical study are twofold. First, it has been shown that ignoring the Doppler shift in such a high-frequency, high-mobility system (modeled in Configuration~$1$) predictably leads to unacceptably low performance and reliability.

At the same time, it has been shown that by (i)~leveraging the predictability of the LEO satellite movement and (ii)~selecting a moderate number of subcarriers, it becomes sufficient to estimate the Doppler shift \emph{just once per data exchange cycle} and then successfully apply low-complexity frequency and phase compensation mechanisms to maintain high performance levels (Configuration~$3$). The study hence serves as an important building block on the way from the theoretical modeling of sub-THz satellite communication links toward their prototyping and possible development for beyond-6G wireless networks. 
The presented framework can be further extended by incorporating 
imperfect beam realignment and adverse weather effects, as well as by considering other sub-THz satellite scenarios, such as inter-satellite links within a LEO constellation.
%The presented framework can also be expanded further to generalize the scenario and conclusions, accounting for, among other factors, realistic imperfect beam realignment strategies and possibly adverse weather conditions, as well as analyzing additional categories of sub-THz satellite communication scenarios, such as sub-THz inter-satellite links within a LEO constellation.

%%%%%%%%%%%%%%%%%%%%%%%%%%%%%%%%%%%%%%%%%%%%%%%%%%%%%%%%%%%%%%%%%%%%%%%%%%%%%%%%%%%%%%%%%%%%%%%%%%%%%%%%%%%%%
\section*{Acknowledgement}
\label{sec:ack}
The work has been supported by the project 6GSTAR by Vinnova, Sweden.

\balance
\bibliographystyle{IEEEtran}
\bibliography{bibl}

@ARTICLE{circular_orbit_table,
  author={Abdelsadek, Mohammed Y. and others},
  journal={IEEE Open Journal of the Communications Society}, 
  title={Broadband Connectivity for Handheld Devices via {LEO} Satellites: {I}s Distributed Massive {MIMO} the Answer?}, 
  year={2023},
  volume={4},
  number={},
  pages={713-726},
  month = {March},
  }

@ARTICLE{truly_mobile_thz,
  author={Petrushkevich, S. and Petrov, V. and Jornet, J.M.},
  journal={NPJ Wireless Technology},
  volume={2},
  number={71},
  title={Truly Mobile Sub-Terahertz Communications Beyond {6G} via Just-in-Time IMU-Assisted Beam Management}, 
  year={2026},
  month={Sept.},
  }

@INPROCEEDINGS{mobile_thz_trial_japan,
  author={Koda, Yusuke and Okamatsu, Muko and Mori, Shota and Ohmi, Norichika and Harada, Hiroshi},
  booktitle={Proc. of the IEEE CSCN}, 
  title={Fundamental Mobile Transmission Trial of Wideband {5G NR} Signal at Sub-{THz} Band in Road Environment}, 
  year={2025},
  volume={},
  number={},
  pages={1-6},
  month={Sept.},
  }

@INPROCEEDINGS{Aliaga2025Scintillation,
  author={Aliaga, Sergi and others},
  booktitle={Proc. of the IEEE CCNC}, 
  title={{Analysis of Scintillation Effects in Terahertz Band Satellite Communications for {6G} and Beyond}}, 
  year={2025},
  volume={},
  number={},
  pages={1-6},
  month={Jan.},
  }

@inproceedings{sen2022terahertz,
  title={Terahertz communications can work in rain and snow: {I}mpact of adverse weather conditions on channels at 140 {GHz}},
  author={Sen, Priyangshu and others},
  booktitle={Proc. of the MmNets wrks. at ACM MobiCom},
  pages={13--18},
  year={2022},
  month={Oct.},
}

@article{ma2018effect,
  title={The effect of snow on a terahertz wireless data link},
  author={Ma, J. and Adelberg, J. and Shrestha, R. and Moeller, L. and Mittleman, D. M.},
  journal={Journal of Infrared, Millimeter, and Terahertz Waves},
  volume={39},
  number={6},
  pages={505--508},
  year={2018},
  publisher={Springer},
  month={June},
}

@ARTICLE{chong_space_sagin,
  author={Han, Chong and Gao, Weijun and Yang, Chuang and Peng, Mugen and Zhang, Wenjun},
  journal={IEEE Wireless Communications}, 
  title={Joint Communication and Radar Sensing for Terahertz Space-Air-Ground Integrated Networks {(SAGIN)}}, 
  year={2026},
  volume={33},
  number={3},
  pages={204-211},
  month={June},
  }

@ARTICLE{60016,
  author={van Exter, M. and Grischkowsky, D.R.},
  journal={IEEE Transactions on Microwave Theory and Techniques}, 
  title={Characterization of an optoelectronic terahertz beam system}, 
  year={1990},
  volume={38},
  number={11},
  pages={1684-1691},
  month={Nov.},
  }

@INPROCEEDINGS{9838922,
  author={Xing, Yunchou and Kanhere, Ojas and Ju, Shihao and Rappaport, Theodore S.},
  booktitle={Proc. of the IEEE ICC}, 
  title={Sub-Terahertz Wireless Coverage Analysis at 142 {GHz} in Urban Microcell}, 
  year={2022},
  volume={},
  number={},
  pages={3942-3947},
  month={May},
  }

@article{sen2023multi,
  title={Multi-kilometre and multi-gigabit-per-second sub-terahertz communications for wireless backhaul applications},
  author={Sen, Priyangshu and Siles, Jose V and Thawdar, Ngwe and Jornet, Josep M},
  journal={Nature Electronics},
  volume={6},
  number={2},
  pages={164--175},
  year={2023},
  month={Feb.},
}

@ARTICLE{11538050,
  author={Nishibori, Toshiyuki and Okada, Nozomi and Kimura, Kimihiro and Sato, Toshio and Tamesue, Kazuhiko and Jitsuno, Kunihisa and Sato, Takuro and Kawanishi, Tetsuya},
  journal={IEEE Transactions on Microwave Theory and Techniques}, 
  title={A 100-{GHz} High-Gain Airborne Antenna System With Ground-to-Air Flight Demonstration Toward {HAPS} Feeder Links}, 
  year={2026},
  volume={74},
  number={7},
  pages={6682-6695},
  month={July},
  }

@misc{ieee_802_15_3d_standard_2024,
author={IEEE},
title="{IEEE Standard for Wireless Multimedia Networks}",
howpublished={IEEE 802.15.3--2023},
month={Feb.},
year={2024},
}

@INPROCEEDINGS{Aliaga2025Payload,
  author={Benincasa, Andrew and others},
  booktitle={Proc. of the Small Satellite Conference}, 
  title={A Radio Payload for In-Space Sub-{T}erahertz Communications}, 
  year={2025},
  pages={1-15},
  month={Aug.},
  }

@INPROCEEDINGS{11309975,
  author={Del Prever, Pietro Brach and others},
  booktitle={Proc. of the IEEE MILCOM}, 
  title={Pointing-Error-Induced Fading in an Open-Loop {THz} Uplink with Hardware Impairments}, 
  year={2025},
  volume={},
  number={},
  pages={245-250},
  month={Oct.},
  }

@article{sat_thz_prototype,
  title={Sub Terahertz {LEO} Satellite Communication: {V}ision, Opportunities, and Challenges toward the First Prototype in Space},
  author={Aliaga, Sergi and others},
  journal={arXiv:2606.15410v2},
  year={2026},
  month={June},
}

@misc{6G_OFDM_agreement,
  author = "{3GPP}",
  title = "{Final Minutes Report, RAN1 \#122 v100}",
  Howpublished =  {R1-2506702},
  year = {2025},
  month = {Sept.},
}

@misc{3GPP38300, 
    author={3GPP},
    title={{NR} and {NG-RAN} {O}verall {D}escription; {S}tage 2},
    Howpublished = {TS 38.300 v.19.3.0},
    year = {2026},
    month = {June},
}

@misc{3gpp38211, 
    author={3GPP},
    title={{NR; Physical Channels and Modulation}},
    Howpublished = {TS 38.211 v.19.4.0},
    year = {2026},
    month = {June},
}

@ARTICLE{mike_thz_spectrum,
  author={Marcus, Michael J.},
  journal={IEEE Wireless Communications}, 
  title={{6G} Spectrum Policy Issues Above 100 {GHz}}, 
  year={2021},
  volume={28},
  number={6},
  pages={7-8},
  month = {Dec.},
  }

@misc{afrl_flight_demo_thz,
author={Denofio, Marc},
title={{AFRL} conducts first flight experiments for communications in terahertz band},
howpublished={U.S. Air Force Research Laboratory},
year={2023},
}

@misc{3GPP38811, 
    author={3GPP},
    title={Study on {N}ew {R}adio {(NR)} to support non-terrestrial networks},
    Howpublished = {TR 38.811 v.15.4.0},
    year = {2020},
    month = {Oct.},
}

@misc{3gpp38214, 
    author={3GPP},
    title={{NR;} Physical Layer Procedures for Data},
    Howpublished = {TS 38.214 v.19.4.0},
    year = {2026},
    month = {June},
}

@article{huang2021synchronization,
  title={Synchronization for {OFDM}-based satellite communication system},
  author={Huang, Miaona and Chen, Jun and Feng, Suili},
  journal={IEEE Transactions on Vehicular Technology},
  volume={70},
  number={6},
  pages={5693--5702},
  year={2021},
  month={June},
}

@ARTICLE{katwe2025cmwave,
  author={Katwe, Mayur V. and others},
  journal={IEEE Wireless Communications}, 
  title={CmWave and Sub-{THz}: {K}ey Radio Enablers and Complementary Spectrum for 6{G}}, 
  year={2025},
  volume={32},
  number={6},
  pages={182-190},
  month={Dec.},
  }

@article{lin2022path,
  author={Lin, Xingqin and Cioni, Stefano and Charbit, Gilles and Chuberre, Nicolas and Hellsten, Sven and Boutillon, Jean-Francois},
  journal={IEEE Communications Magazine}, 
  title={On the Path to {6G}: {E}mbracing the Next Wave of Low {E}arth Orbit Satellite Access}, 
  year={2021},
  volume={59},
  number={12},
  pages={36-42},
  month={Dec.},
  }

@article{giordani2020non,
  author={Giordani, Marco and Zorzi, Michele},
  journal={IEEE Network}, 
  title={Non-Terrestrial Networks in the {6G} Era: {C}hallenges and Opportunities}, 
  year={2021},
  volume={35},
  number={2},
  pages={244-251},
  month={Dec.},
  }

@article{zuo2023ofdm,
  title={{OFDM}-based massive connectivity for {LEO} satellite {I}nternet of {T}hings},
  author={Zuo, Yong and Yue, Mingyang and Zhang, Mingchen and Li, Sixian and Ni, Shaojie and Yuan, Xiaojun},
  journal={IEEE Transactions on Wireless Communications},
  volume={22},
  number={11},
  pages={8244--8258},
  year={2023},
  month={Nov.},
}

@article{na2013channel,
  title={Channel model and throughput analysis for {LEO OFDM} satellite communication system},
  author={Na, Zhenyu and Guan, Qinyang and Fu, Ce and Cui, Yang and Guo, Qing},
  journal={International Journal of Future Generation Communication and Networking},
  volume={6},
  number={6},
  pages={109--122},
  year={2013},
  month={June},
}

@article{hassan2024spaceris,
  title={SpaceRIS: {LEO} satellite coverage maximization in {6G} sub-{THz} networks by {MAPPO DRL} and whale optimization},
  author={Hassan, Sheikh Salman and Park, Yu Min and Tun, Yan Kyaw and Saad, Walid and Han, Zhu and Hong, Choong Seon},
  journal={IEEE Journal on Selected Areas in Communications},
  volume={42},
  number={5},
  pages={1262--1278},
  year={2024},
  month={May},
}

@article{bakhsh2025direct,
  title={Direct uplink connectivity in space {MIMO} systems with {THz} and {FSO} intersatellite links},
  author={Bakhsh, Zohre Mashayekh and Omid, Yasaman and Chen, Gaojie and Ma, Yi and Kayhan, Farbod and Tafazolli, Rahim},
  journal={IEEE Systems Journal},
  volume={19},
  number={4},
  pages={1270--1281},
  year={2025},
  month={Dec.},
}

@article{armstrong1999analysis,
  title={Analysis of new and existing methods of reducing intercarrier interference due to carrier frequency offset in {OFDM}},
  author={Armstrong, Jean},
  journal={IEEE Transactions on Communications},
  volume={47},
  number={3},
  pages={365--369},
  year={1999},
  month={Mar.},
}

@article{zhao2001intercarrier,
  title={Intercarrier interference self-cancellation scheme for {OFDM} mobile communication systems},
  author={Zhao, Yuping and Haggman, S-G},
  journal={IEEE Transactions on Communications},
  volume={49},
  number={7},
  pages={1185--1191},
  year={2001},
  month={July},
}

@article{pollet1995ber,
  title={{BER} sensitivity of {OFDM} systems to carrier frequency offset and {Wiener} phase noise},
  author={Pollet, Thierry and Van Bladel, Mark and Moeneclaey, Marc},
  journal={IEEE Transactions on Communications},
  volume={43},
  number={2/3/4},
  pages={191--193},
  year={1995},
  month={Feb./Mar./Apr.},
}

@inproceedings{kim2007residual,
  title={Residual frequency offset and phase compensation for {OFDM} systems},
  author={Kim, Wonchae and Cox, Donald C},
  booktitle={Proc. of the IEEE VTC},
  pages={2209--2213},
  year={2007},
  month={Sept.-Oct.}
}

@INPROCEEDINGS{HenkelCyclicPrefix2002,
  author={Henkel, W. and Taubock, G. and Odling, P. and Borjesson, P.O. and Petersson, N.},
  booktitle={Proc. of the International Zurich Seminar on Broadband Communications Access -- Transmission -- Networking}, 
  title={The cyclic prefix of {OFDM/DMT} - {A}n analysis}, 
  year={2002},
  volume={},
  number={},
  pages={22-22},
  month={Feb.},
  }

@inproceedings{batariere2004cyclic,
  title={Cyclic prefix length analysis for {4G} {OFDM} systems},
  author={Batariere, Mickael and Baum, Kevin and Krauss, Thomas P},
  booktitle={Proc. of the IEEE VTC},
  volume={1},
  pages={543--547},
  year={2004},
  month={Sept.},
}

@ARTICLE{cooper2025power,
  author={Cooper, Ken B. and Siles, Jose V. and Socuellamos, Juan M. and Lee, Choonsup and Lin, Robert H. and Escobar, Uriel S. and Dengler, Robert J. and Monje, Raquel Rodriguez},
  journal={IEEE Journal of Microwaves}, 
  title={A Power-Combined 240 {GHz} Frequency-Multiplier Source for Cloud Radar Applications}, 
  year={2025},
  volume={5},
  number={6},
  pages={1253-1259},
  month={Nov.},
  }

@ARTICLE{Ali1998Doppler,
  author={Ali, I. and Al-Dhahir, N. and Hershey, J.E.},
  journal={IEEE Transactions on Communications}, 
  title={Doppler characterization for {LEO} satellites}, 
  year={1998},
  volume={46},
  number={3},
  pages={309-313},
  month={Mar.},
  }

@book{casella2024statistical,
  title={Statistical inference},
  author={Casella, George and Berger, Roger},
  year={2024},
  publisher={Chapman and Hall/CRC}
}

@article{simon1998some,
  title={Some new twists to problems involving the {Gaussian} probability integral},
  author={Simon, Marvin K and Divsalar, Dariush},
  journal={IEEE Transactions on Communications},
  volume={46},
  number={2},
  pages={200--210},
  year={1998},
  publisher={IEEE}
}

@book{maral2020satellite,
  title={Satellite communications systems: systems, techniques and technology},
  author={Maral, G{\'e}rard and Bousquet, Michel and Sun, Zhili},
  year={2020},
  publisher={John Wiley \& Sons}
}

@MISC{fcc2022grantSpaceX,
  Author = {FCC},
  Title = {{FCC} Partially Grants {SpaceX Gen2} Broadband Satellite Application},
          month = {Nov.},
        year = {2022},
}

@techreport{ITU_M2541_2024,
  author      = {{ITU-R}},
  title       = {Technical Feasibility of {IMT} in Bands Above 100 {GHz}},
  institution = {International Telecommunication Union Radiocommunication Sector (ITU-R)},
  type        = {Report ITU-R M.2541-0},
  year        = {2024},
  month       = may,
}

@book{proakis2001digital,
  title={Digital communications},
  author={Proakis, John G and Salehi, Masoud},
  volume={4},
  year={2001},
  publisher={McGraw-hill New York}
}

@ARTICLE{tanash2026performance,
  author={Tanash, Islam M. and González-Prelcic, Nuria and Wichman, Risto},
  journal={IEEE Wireless Communications Letters}, 
  title={Performance Analysis of {LEO}-Terrestrial Systems in Presence of {Doppler} Effect}, 
  year={2026},
  volume={15},
  number={},
  pages={4205-4209},
  month={July},
  }

@techreport{3GPP38821,
    author      = {{3GPP}},
    title       = {{Solutions for NR to Support Non-Terrestrial Networks (NTN)}},
    institution = {{3rd Generation Partnership Project (3GPP)}},
    number      = {TR 38.821},
    type        = {Technical Report},
    year        = {2023}
}

\vfill

\end{document}